\newcommand{\challenge}{PROCESS}
\documentclass{article}
\usepackage{spconf,amsmath,graphicx}
\usepackage{xcolor}
\usepackage{hyperref}
\usepackage{multirow}
\usepackage{booktabs} 

\title{Early Dementia Detection Using Multiple Spontaneous Speech Prompts: The PROCESS Challenge}
\name{
    \parbox{\textwidth}{
        \centering
        \textit{Fuxiang Tao$^1$, Bahman Mirheidari$^1$, Madhurananda Pahar$^1$, Sophie Young$^1$, Yao Xiao$^1$, }\\    
        \textit{Hend Elghazaly$^1$, Fritz Peters$^1$, Caitlin Illingworth$^1$,  Dorota Braun$^1$, Ronan O'Malley$^1$, Simon Bell$^1$, Daniel Blackburn$^1$, Fasih Haider$^2$, Saturnino Luz$^2$, Heidi Christensen$^1$}\\
        }
}

\address{$^1$University of Sheffield, UK\\ $^2$University of Edinburgh, UK}

\begin{document}
\maketitle
\begin{abstract}
Dementia is associated with various cognitive impairments and typically manifests only after significant progression, making intervention at this stage often ineffective. To address this issue, the Prediction and Recognition of Cognitive Decline through Spontaneous Speech (PROCESS) Signal Processing Grand Challenge invites participants to focus on early-stage dementia detection. We provide a new spontaneous speech corpus for this challenge. This corpus includes answers from three prompts designed by neurologists to better capture the cognition of speakers. Our baseline models achieved an F1-score of 55.0\% on the classification task and an RMSE of 2.98 on the regression task.
\end{abstract}

\section{Introduction}
%
Dementia is an umbrella term that describes symptoms arising from several conditions affecting individuals' memory, speech, communication, cognitive abilities, and daily functioning. 
Dementia is typically recognized only when symptoms become pronounced; however, its effects on the brain begin much earlier. Mild cognitive impairment (MCI), often regarded as a transitional stage between normal ageing and dementia, involves mild declines in cognitive functions, such as memory and attention, without significantly disrupting daily activities. While timely diagnosis of MCI could lead to the implementation of effective interventions and treatments, its detection remains challenging.

In recent years, there have been a number of challenges designed to attract interest in this field of study~\cite{luz2024overview}. 
Most of these were developed using data sets available on Dementia Bank. While those data provided useful initial benchmarks, new data is needed for a number of reasons.
First, patient voice samples in Dementia Bank come mostly from Alzheimer's dementia patients at late stages of the disease, making the task of distinguishing between patients and controls relatively easy both from a clinical and from a machine learning perspective. Second, the audio quality of the data is poor and does not represent the quality that it is possible to achieve even with current, standard consumer-based devices like modern laptops. These factors underscore the necessity for new data sets to ensure the continued advancement and accuracy of research in this field. 

The \challenge{} Signal Processing Grand Challenge aims to establish a platform for contributions and discussions on early-stage dementia detection using speech signal processing and Artificial Intelligence (AI) models. To support this, we provide a state-of-the-art corpus covering a broader range of diagnostic classes for different subtypes of early-stage dementia, including mild cognitive impairment (MCI). We invite paper submissions presenting innovative AI models, novel speech signal processing techniques, and advanced feature selection and extraction methods for the \challenge{}.
\section{The Tasks and the Corpus}
The \challenge{} Signal Processing Grand Challenge consists of two tasks: 1) A classification task that is aimed at distinguishing early cognitive decline and dementia from healthy volunteers via speech; 2) A regression task that predicts the Mini Mental State Examination (MMSE) scores from speech. Challenge participants could choose to undertake one or both tasks and the evaluation criteria is based on their choice. 
Participants received training and development sets to train their models, along with an independent test set provided one week before the results submission deadline. Participants were allowed to submit up to 3 results, from 3 different models, and the best model was considered. Invited papers describe the models they attempted.
%

The corpus used in the challenge was collected via an online platform (\url{https://www.cognospeak.com/}). The corpus consists of speech elicited by three different tasks that were designed based on neuroscience research for dementia diagnosis: the \emph{Semantic Fluency}, the \emph{Phonemic Fluency}, and the \emph{Cookie Theft} picture description. All audio was spoken in English. \emph{Semantic Fluency} asks participants to respond to the task of ``Please name as many animals as you can in a minute.'' This is similar to the naming task in many cognitive assessments, which primarily evaluate language abilities and naming skills to detect potential issues in language comprehension and expression. \emph{Phonemic Fluency} asks participants to respond to the task of ``Please say as many words beginning with the letter `P' as you can. Any word beginning with `P' except for names of people such as Peter, or countries such as Portugal.'' The time limit for this is also a minute and this task is similar to the language tasks used in cognitive assessments to test verbal fluency and executive functions related to language. \emph{Cookie Theft} is a picture description task that is widely used to collect audio in dementia detection research. This type of audio is expected to reflect various cognitive functions of speakers, such as language comprehension and memory.

The training and development sets include audio recordings and manual transcripts for each speaker's prompt to train Automatic Speech recognition (ASR) models. Manual transcripts are excluded from the test set. Speaker diagnoses (healthy, MCI, dementia) are provided for classification, and MMSE scores for regression.

\textbf{Evaluation.}\label{sec:eva}
For the classification task, the evaluation metrics include macro accuracy, macro precision, macro recall, and macro F1-score. For the regression task, the evaluation metric is root mean squared error (RMSE). The rankings of participants are based on F1-score and RMSE scores.


\section{Baseline Models}
We present baseline models using acoustic and text features. Acoustic features were extracted with OpenSmile~\cite{Eyben2013} and eGeMAPS feature set~\cite{eyben2015geneva}, converting recordings into a sequence of vectors $X=\{\vec x_1, \vec x_2, \ldots, \vec x_T\}$, where $T$ is the total number of vectors averaged for each prompt. These were input into a Support Vector Machine (SVC) with a linear kernel and Random Forest Classification (RFC). Table~\ref{tab:classification} shows baseline results, with the best F1-score of 55.0\% for VF (semantic + phonemic fluency). For regression, average vectors 
$X$ were input into Support Vector Regression (SVR) and Random Forest Regression (RFR). Table~\ref{tab:regression} reports RMSEs of 4.40 and 3.30 for CT; 7.93 and 3.31 for VF; and 6.68 and 3.17 for their combination. For the text baseline model, speech recordings were transcribed using Whisper ~\cite{radford2023robust}, an ASR system, and the transcripts were processed by RoBERTa~\cite{liu2019roberta}. Table~\ref{tab:classification} reports the best F1-score of 36.8\%. For regression, the transcripts were input into RoBERTa, yielding RMSEs of 2.99 for CT, 2.98 for both VF and CT+VF (Table~\ref{tab:regression}). 
\begin{table}[ht]

\centering
\caption{Baseline classification results for the cookie theft (CT), semantic fluency and phonemic fluency (VF) and their combination (CT+VF) are presented in terms of accuracy (Acc.), precision (Prec.), recall (Rec.), and F1-score, reported as percentages.}\label{tab:classification} 
\begin{tabular}{p{1.6cm}p{0.95cm}p{0.6cm}p{0.6cm}p{0.6cm}p{0.6cm}}\toprule
                                       Model   &  Prompt & Acc. & Prec. & Rec. & F1 \\ \midrule
\multicolumn{1}{l|}{\multirow{3}{*}{SVC (eGeMAPS)}} & CT & 57.5 &  53.5 & \textbf{61.2} & \textbf{55.0}\\
\multicolumn{1}{l}{}                     & VF & 45.0 &  38.8 & 39.1 & 38.3\\
\multicolumn{1}{l|}{}                     & CT+VF & 50.0 &  41.7 & 42.3 & 41.7\\ \midrule
\multicolumn{1}{l|}{\multirow{3}{*}{RFC (eGeMAPS)}}  & CT & \textbf{60.0} &  \textbf{71.7} & 49.9 & 53.3\\
\multicolumn{1}{l|}{}                     & VF & 52.5 &  33.1 & 35.9 & 33.9\\
\multicolumn{1}{l|}{}                     & CT+VF & 52.5 &  66.1 & 43.9 & 47.4\\ \midrule
\multicolumn{1}{l|}{\multirow{3}{*}{RoBERTa-Classifier}}  & CT & 52.5 & 36.1 & 39.7 & 36.8\\
\multicolumn{1}{l|}{}                     & VF & 55.0 & 35.6 & 38.1 & 35.6\\
\multicolumn{1}{l|}{}                     & CT+VF & 52.5 & 32.2 & 35.9 & 32.9\\ \bottomrule
\end{tabular}
\vspace{-0.5cm}
\end{table}


\begin{table}[ht]
\centering
\caption{Baseline regression results are presented in terms of root mean squared error (RMSE).}\label{tab:regression} 
\begin{tabular}{cccc}\toprule
                                    Model      & CT & VF & CT+VF\\ \midrule
\multicolumn{1}{c|}{{SVR (eGeMAPS)}} & 4.40 & 7.93 & 6.68 \\ \midrule
\multicolumn{1}{c|}{{RFR (eGeMAPS)}}  & 3.31 & 3.31  & 3.17 \\ \midrule
\multicolumn{1}{c|}{{RoBERTa-Regression}}  & 2.99 & \textbf{2.98} &\textbf{2.98} \\ \bottomrule
\end{tabular}
\vspace{-1em}
\end{table}


\section{Conclusion}
Our work suggests that utilizing acoustic and textual features from spontaneous speech shows promise for early-stage dementia detection, aiding both in identifying MCI/dementia and in estimating the severity of cognitive impairment. However, the performances of current baseline models highlight a significant need for improvement.
We hope this work can serve as a benchmark for evaluating early-stage dementia detection and encourage more work to contribute to this topic.

\bibliographystyle{IEEEbst}
\bibliography{proposal}

\end{document}